\documentclass[letterpaper,twocolumn,10pt]{article}
\usepackage{usenix2019_v3}

\usepackage{tikz}
\usepackage{amsmath}

\usepackage[labelfont=bf]{caption}
\usepackage{subcaption}
\usepackage{hyperref}

\newcommand{\ignore}[1]{}
\newcommand{\high}[1]{\textcolor{blue}{#1}}


\newcommand{\appref}[1]{\hyperref[#1]{Appendix~\ref*{#1}}}

\newcommand{\nametwo}{HGRD}

\newcommand{\name}{\nametwo{}}                
\newcommand{\fullname}{\underline{H}ost code \underline{G}uided GPU \underline{R}ace \underline{D}etector}
\newcommand{\racetool}{{\name{}}}                 
\newcommand{\gverify}{GPUVerify}
\newcommand{\faial}{FaialAA}

\newcommand{\gpumc}{GPUMC}
\newcommand{\iguard}{iGUARD}
\newcommand{\gklee}{GKLEE}

\newcommand{\hirace}{HiRace}

\usepackage{tikz}

\usepackage{tcolorbox}

\usepackage[frozencache,cachedir=.]{minted}
\setminted{linenos=true,autogobble,xleftmargin=12pt,numbersep=7pt}
\newenvironment{code}
{\minted[fontfamily=cmtt,escapeinside=@@,fontsize=\scriptsize]{cuda}}
{\endminted}

\newenvironment{pycode}
{\minted[fontfamily=cmtt,escapeinside=@@,fontsize=\scriptsize]{python}}
{\endminted}

\newcommand{\textcode}[1]{{\small \texttt{#1}}}

\usepackage{tikz}
\newcommand{\circled}[1]{\tikz[baseline=(char.base)]{
            \node[shape=circle,draw,inner sep=0.5pt] (char) {#1};}}
\newcommand{\mycircled}[1]{\circled{\textbf{#1}}}
\newcommand{\myparagraph}[1]{\noindent\textbf{\textit{#1}:}}

\newcommand{\ie}{\textit{i.e.,}}
\newcommand{\eg}{\textit{e.g.,}}

\newcommand{\cuda}{CUDA}

\newenvironment{mybullet}
{\begin{list}{$\bullet$}
{\setlength{\topsep}{0mm}\setlength{\itemsep}{0mm}
\setlength{\parsep}{0mm}
\setlength{\listparindent}{\parindent} 
\setlength{\itemindent}{0mm}\setlength{\partopsep}{0mm}
\setlength{\labelwidth}{-2mm}
\setlength{\leftmargin}{0mm}}}
{\end{list}}

\newcommand{\mysection}[1]{\section{#1}}
\newcommand{\mysubsection}[1]{\subsection{#1}}

\usepackage{multirow}

\newcommand{\indigoa}{pathCompression}
\newcommand{\indigob}{pushNode}
\newcommand{\indigoc}{condNeighbor}

\newcommand{\kaldia}{copyUppToLow}
\newcommand{\kaldib}{copyLowToUpp}
\newcommand{\kaldic}{copyFromTp}
\newcommand{\reduction}{reduction}
\newcommand{\expdist}{expDistance}
\newcommand{\random}{randomAccess}
\newcommand{\tone}{toneMapping}
\newcommand{\convSep}{convSeparable}
\newcommand{\lud}{luDecomposition}
\newcommand{\nw}{needlemanWunsch}
\newcommand{\texBind}{pitchLinear}
\newcommand{\copyRows}{copyFromVec}

\newcommand{\access}{\textsf{access}}
\newcommand{\accesspair}{\textsf{access} \textsf{pair}}

\usepackage[framemethod=tikz]{mdframed}
\newcommand{\inlinerounded}[1]{%
  \tikz[baseline=(X.base)] \node[draw,rounded corners=2pt,inner xsep=1.5pt,inner ysep=1.5pt] (X) {#1};}
\newcommand{\component}[1]{\inlinerounded{\textbf{C#1}}}

\newcommand{\codebox}[1]{#1}

\begin{document}




\author{
{\rm Ajay Nayak}\\
Indian Institute of Science
\and
{\rm Anubhab Ghosh}\\
Indian Institute of Science
\and
{\rm Arkaprava Basu}\\
Indian Institute of Science
} 


\title{\Large \bf Towards an Accurate GPU Data Race Detector}

\maketitle

\begin{abstract}

Data races in GPU programs pose a threat to the reliability of GPU-accelerated software stacks.
Prior works proposed various dynamic (runtime) and static (compile-time) techniques to detect races in GPU programs. 
However, dynamic techniques often miss critical races, as they require the races to manifest during testing. 
While static ones can catch such races, they often generate numerous false alarms by conservatively assuming values of variables/parameters that cannot ever occur during any execution of the program. 

We make a key observation that the host (CPU) code that launches GPU kernels contains crucial semantic information about the values that the GPU kernel's parameters can take during execution.
Harnessing this hitherto overlooked information helps accurately detect data races in GPU kernel code. 
We create \racetool{}, a new state-of-the-art static analysis technique that performs a holistic analysis of \textit{both} CPU and GPU code to accurately detect a broad set of true races while minimizing false alarms. 
While SOTA dynamic techniques, such as \iguard{}, miss many true races, \racetool{} misses none. 
On the other hand, static techniques such as \gverify{} and \faial{} raise tens of false alarms, where \racetool{} raises none.

\end{abstract}





\mysection{Introduction}
\label{sec:intro}

Modern software, from AI and HPC to graph processing, relies on Graphics Processing Units (GPUs) for its primary computing needs~\cite{google-cloud-gpu,aws-gpu,azure-gpu:nvidia}.
Consequently, bugs (errors) in GPU-accelerated programs can significantly impact the reliability of a large section of today's software ecosystem.

Data races in GPU programs (kernels) pose a critical threat to the reliability of GPU-accelerated programs~\cite{iGUARD,Simulee,BARRACUDA}. 
These are synchronization bugs that occur when two or more threads access shared data with at least one of them writing it, and the accesses are not separated by synchronization.
Races may manifest \textit{only} under specific inputs and/or under certain non-deterministic events (\eg{} order of loads and stores). 

Even when a data race exists in a buggy program, it \emph{may not} manifest in many or most of its executions, and thus can escape software testing.
However, when races manifest in software during deployment, it can crash the program or produce erroneous results~\cite{shanlu-multicore,ceze-lard,conseq,lu-transac}. 

Detecting data races in GPU programs is challenging due to their massive parallelism, hierarchical programming, and a synchronization model that is unique to GPUs~\cite{iGUARD,ScoRD}. 
Thus, several recent research works have proposed various techniques to detect races in GPU programs ~\cite{iGUARD,GKLEE,CURD,BARRACUDA,GRace,GMRace,LDetector,PUG,ScoRD,HAccRG,Faial,Simulee}.  
These broadly fall into two categories: 
\mycircled{1} dynamic techniques that trace memory and synchronization accesses \textit{at runtime} to report races~\cite{iGUARD,HAccRG,ScoRD,HiRace,BARRACUDA,CURD}, and 
\mycircled{2} static ones that report races through source code analysis while \textit{compiling} the GPU program~\cite{gpuverify,PUG,gpumc,Faial,dartagnan}. 
However, both suffer from critical shortcomings in their current incarnations.

Dynamic tools can detect races during an execution of a program \emph{only if the races manifest during that execution}~\cite{eraser,fasttrack}.
They can \textit{fail} to detect races even if race(s) exist in a program and are fundamentally prone to miss races that manifest infrequently.
Unfortunately, such races are more likely to escape testing and pose a critical reliability threat to software in deployment. 
In addition, these tools must trace memory accesses and synchronization operations during execution, adding significant runtime (\eg{} $60\times$ for \iguard{}~\cite{iGUARD}) and memory overheads (\eg{} at least $4\times$ for \iguard{}). 
Such overheads prevent them from deployment in production, limiting their usefulness to smaller programs with limited inputs.

In contrast, static (compile-time) analysis of source code can detect the presence of data races that may \textit{not} have been manifested in any execution yet~\cite{gpuverify,Faial}.
It can catch infrequently manifesting critical races. 
Unfortunately, these techniques are prone to reporting false positives, \ie{} false alarms of races when none exist. 

In the absence of runtime information on concrete values of the program's variables and parameters, static techniques must analyze programs considering \textit{all possible} values the variables (parameters) can assume during any execution.
Unfortunately, it is common to conservatively consider values of a program's variables/parameters that can \textit{never} occur in any execution in practice.
As a consequence, they report false positives. 
The usefulness of a race detector quickly diminishes with an increasing propensity to report false positives. 
Thus, a key challenge is to accurately estimate the set of possible values of variables and parameters -- the primary focus of this work. 

\myparagraph{A case for host code guided GPU race detection} 
Our key insight is that crucial \emph{semantic information} about the values that the variables and parameters in a GPU kernel can assume during execution is often \emph{embedded in the associated host (CPU) code}. 
A GPU-accelerated program consists of one or more GPU kernels and the host code which allocates memory for data structures, sets the arguments to pass to the kernels, specifies the number of GPU threads (thread grid dimension) to spawn, and launches the kernels. 

In hindsight, it is no surprise that valuable information can be gleaned from the host code to \textit{accurately} detect races in GPU kernels.
An accurate race detector must report true races and minimize false positives.   
Unfortunately, existing static techniques are myopic, \ie{} they analyze \textit{only} the GPU kernels to detect races in them. 
We discovered \textit{five classes} of semantic information in the host code of real-world programs that can help accurately detect races: 

\mycircled{1} Programmers often embed constraints on the kernel parameters as `asserts' in the host code. 
For example, programs from Kaldi~\cite{Kaldi}, a popular toolkit for speech recognition, launch certain GPU kernels only for square matrices (\ie{} the number of rows and columns are equal).
Ignoring this information could make static analysis techniques report potential races in the GPU kernel. 
However, such a race can \textit{never} happen in \textit{any execution} of the program. 

\mycircled{2} At times, the host code launches GPU kernels with \textit{one} group of related GPU threads (threadblock in CUDA) for certain computations. 
For example, while performing a parallel reduction~\cite{CUDASamples} for a large array over multiple rounds in a hierarchical fashion, a single threadblock is tasked with computing the final result in the \textit{last} round.
Ignoring this information in the host code would lead to \textit{false alarms} that threads from concurrent threadblocks can race.

\mycircled{3} We observed that different kernel parameters can be semantically related. 
For example, it is typical for host code to set the number of GPU threads (thread grid dimension) based on the dimensions of an input image or matrix that the kernel would compute. 
Such relationships among seemingly independent parameters are \textit{not} discernible by analyzing the kernel code but are easily fathomable from the host code.

\mycircled{4} The range of values that a kernel parameter may assume can be limited by a loop (\eg{} \textcode{for}) iterator variable in the host code. 
For example, it is typical for the host code to iteratively launch a kernel to compute on distinct parts (offsets) of a large matrix or image. 
The offset parameter passed to the kernel is limited by the bounds of the loop iterator variable in the host code (\eg{} the dimension of the matrix it must process). 

\mycircled{5} Often, the variables passed to kernels during launch are \textit{also} used to determine the size of data structures (memory allocations). 
For example, the host code typically sets the sizes based on the problem (\eg{} input matrix size). 
The variables that set the size \textit{must} have a \textit{positive} value. 
These variable(s) are often passed as kernel parameter(s) to bound the computations in the kernel code. 
The values of such parameters are \textit{implicitly} limited to only positive values due to their use in sizing memory allocations. 

\myparagraph{Key observation} 
\emph{We identified five classes of semantic information embedded in the \underline{host code} that limit the set of values that variables in the kernel code can assume during any execution.
It is imperative to leverage this knowledge to accurately detect \underline{data races in GPU kernels}}. 

\myparagraph{Broadening the ambit of race detection} 
Further, we observe that existing static analyses cannot detect data races due to advanced synchronization features in GPUs, \eg{} intra-warp races (\autoref{sec:background}). 
Many are oblivious to fine-grain acquire-release synchronization (\eg{} \gverify{}~\cite{gpuverify}, \faial{}~\cite{Faialaa}) and/or require invasive rewriting of programs (\eg{} \gpumc{}~\cite{gpumc}).
These fundamental shortcomings limit the usefulness of existing techniques, especially with the growing use of fine-grain synchronization for improving utilization~\cite{scopeadvice,ScoRD}. 
A deeper analysis of the kernel code can circumvent these limitations.
For example, it is possible to accurately detect races in the presence of lock acquire/release by relating the acquire-release to the data accesses that it intend to guard. 

\myparagraph{\racetool{}}
We create a new state-of-the-art static analysis tool called \racetool{} (\fullname{}) to detect data races in GPU kernels. 
It is the \emph{first to leverage a holistic analysis of host (CPU) and kernel (GPU) code to accurately detect true races while minimizing false positives.} 
It also broadens the ambit of types of races detected in GPU kernels by current static techniques.
It detects intra-warp data races and races in the presence of acquire-release synchronization. 
Being a static technique, it is \textit{not} limited by the shortcomings of dynamic techniques, \ie{} it reliably catches hard-to-find data races that manifest infrequently and adds no runtime overheads. 

On a diverse set of $22$ GPU-accelerated programs~\cite{Kaldi,HeCBench,gpuverify,CUDASamples,rodinia,scor}, \racetool{} accurately detected fifteen true data races \textit{without} any false positives. 
Dynamic analysis tools like \iguard{}~\cite{iGUARD} failed to detect nine races as they did not manifest during execution. 
On the other hand, up to $\sim$$79\%$ of the reported races by static analysis tools like \gverify{}~\cite{gpuverify} and \faial{}~\cite{Faialaa} were false positives. 

\noindent\textbf{\underline{Contributions:}} 
\racetool{} is the new state-of-the-art for detecting data races in GPU-accelerated programs that uses static analysis.
It makes two primary contributions: 
\mycircled{1} It is the first to demonstrate the need to analyze host code to glean crucial semantic information for accurate race detection in the GPU kernel code. 
\mycircled{2} It broadens the types of fine-grain data races, \eg{} intra-warp race, that no current static analysis detects. 

\mysection{Background}
\label{sec:background}


To support thousands of concurrent threads, GPU hardware resources are organized in a hierarchy. 
The basic execution block on GPUs is a Streaming Multiprocessor (SM) consisting of multiple Single Instruction Multiple Data (SIMD) units and a scratchpad (shared memory). 
SIMD units have multiple execution lanes (\eg{} $32$) that execute the same instruction on different data items in parallel. 
Threads executing on the same SM can efficiently communicate via the SM's scratchpad, which is typically tens of KBs in size. 
All GPU threads can access and communicate via the global memory, which consists of the GPU's onboard GDDR or HBM memory. 
It is typically tens to a couple of hundreds of GBs in size. 

The GPU programming model mimics the underlying hardware, with the smallest unit of execution being a thread that occupies a single SIMD lane. 
A group of threads (\eg{} $32$) form a warp that can be scheduled on a SIMD unit and typically executes in lockstep. 
A threadblock is a collection of warps that reside on the \textit{same} SM and can communicate among themselves via shared memory. 
A grid consists of all the threadblocks used to execute a GPU kernel. 

\begin{mybullet}

\item \textbf{GPU kernel launch and kernel parameters:}
GPU-accelerated programs consist of GPU kernels that execute on the GPU, and a CPU program (host code) responsible for the overall execution, including launching these GPU kernels. 

The GPU kernel contains instructions that \textit{each} GPU thread should execute, while the host code launches the kernel with thousands of such GPU threads (called a thread grid). 
The host code sets up thread grid dimensions using 3-dimensional built-in variables \textcode{gridDim} and \textcode{blockDim}. 
Dimensions from these variables are accessed by the x, y, and z axes. 
The host code specifies sizes of the thread grid (through \textcode{gridDim}) and the threadblock (through \textcode{blockDim}) while launching a GPU kernel for execution.
For example, launching a kernel with \textcode{gridDim.x} as $2$, and \textcode{blockDim.x} as $128$ creates a thread grid with $2$ threadblocks, each with $128$ threads, \ie{} a total of $256$ GPU threads. 

Each thread in the launched grid has corresponding 3D built-in variables \textcode{threadIdx} and \textcode{blockIdx} with x, y, and z axes. 
A thread is uniquely identified using the values of these variables. 
For example, in a 1-dimensional grid (y and z dimensions of \textcode{gridDim} and \textcode{blockDim} is $1$), a threadblock is uniquely identified by the value of its \textcode{blockIdx.x}. 
The value of `\textcode{blockDim.x} * \textcode{blockIdx.x} + \textcode{threadIdx.x}' uniquely identifies a thread. 
This method to identify threads and threadblocks can be extended to 2 and 3-dimensional grids. 

The host code is also responsible for allocating memory (\eg{} using \textcode{cudaMalloc}) and setting up kernel arguments. 
This includes pointers to data structures and scalar values. 

\item \textbf{Synchronizations:}
Traditionally, GPU programs relied on bulk-synchronous parallelism. 
For example, a block barrier (\textcode{syncthreads}) synchronizes all threads of a threadblock.
However, emerging programs (\eg{} graph processing) require fine-grain synchronization using fences and atomic operations. 
A fence (\eg{} \textcode{threadfence}) ensures that the calling thread is blocked until its writes are visible to other threads. 
Atomic operations (\eg{} \textcode{atomicCAS}) perform read-modify-write updates on an address. 
These primitives allow programmers to build acquire-release synchronization patterns (\eg{} locks). 
An acquire consists of an \textcode{atomicCAS} followed by a \textcode{fence}, while a release consists of a \textcode{fence} followed by an \textcode{atomicExch}~\cite{CUDAProgrammingGuide}.  

Traditionally, threads within a warp executed in lockstep, \ie{} there is an implicit barrier across threads of a warp after every instruction. 
However, since NVIDIA's Volta architecture, a hardware feature called Independent Thread Scheduling can cause threads in a warp to diverge from their lockstep~\cite{nvidia-volta}. 
This feature was introduced to avoid possible deadlocks due to the lockstep execution in programs that use fine-grain locking. 
\cuda{} also introduced warp barrier (\textcode{syncwarp}) for programmers to synchronize threads within a warp when needed~\cite{nvidia-volta}.

\item \textbf{Data races on GPUs:}
A data race occurs when memory accesses from two threads alias to the same address, at least one is a write, and they are not synchronized. 
Data races on GPUs can occur in shared (among threads of a threadblock) and/or global memory (across any thread on the GPU). 

Many prior works efficiently detect races that occur within the threads of a threadblock via the shared memory~\cite{boyer2008ada,CUDARacecheck,GRace,GMRace}. 
However, they ignore the races that can occur across all the threads on the GPU via the global memory. 
Thus, we focus on the challenging global memory races due to the sheer scale (\eg{} number of threads involved, size of the global memory) at which they can occur. 

\myparagraph{Types of data races in GPU kernels}
Three types of races can occur in global memory: inter-, intra-threadblock, and intra-warp. 
An inter-threadblock race occurs when the threads involved belong to different threadblocks. 
An intra-threadblock race occurs when the threads involved belong to the same threadblock but different warps. 
Finally, an intra-warp race occurs when the threads involved belong to the same warp. 

\end{mybullet} 

\section{Prior Works and \racetool{}'s Goals}
\label{sec:prior-goals}

Prior works have proposed both dynamic (runtime) and static (compile-time) techniques to detect races in GPU programs. 
We discuss the primary challenges in these techniques to put the goals of our work in the right context. 

\myparagraph{Dynamic analysis}
Prior works~\cite{BARRACUDA,CURD,HiRace,iGUARD} have used instrumentation based methods to detect races at runtime, \ie{} dynamic analysis. 
A key advantage of this technique is that it reports only true races, \ie{} no false positives. 
However, its fundamental limitation is that it \emph{reports races only if they manifest during execution}. 
State-of-the-art dynamic techniques~\cite{iGUARD,HiRace} fail to report races in programs with true races, \ie{} false negatives (\autoref{sec:eval:analysis}). 
This limits the usefulness of race detectors since most hard-to-find races manifest infrequently. 
Furthermore, dynamic analysis incurs significant runtime overhead (up to $60\times$ for \iguard{}~\cite{iGUARD}) and memory overhead (at least $4\times$ for \iguard{}). 

\myparagraph{Static analysis}
Prior works~\cite{gpuverify,Faial,PUG,gpumc,Faialaa} have proposed analyzing source code to detect races without executing the program, \ie{} static analysis. 
The primary advantage of this technique is that it can detect races that are \textit{yet to manifest} during execution (\eg{} testing). 
This attribute is highly desirable for detecting races as such bugs manifest under specific inputs and/or non-deterministic ordering of memory events~\cite{eraser,fasttrack}. 
Additionally, they do not incur runtime overheads. 

However, static analysis must consider all possible input values a program \textit{may} witness during execution. 
Thus, a key limitation is their \emph{propensity to report false positives}.
It occurs due to the conservative estimation of variable values that can never happen in any execution. 
A race detector that raises many false positives loses its worth. 
Existing techniques also miss different types of races (\eg{} intra-warp races~\cite{gpumc,PUG,Faialaa}, or are oblivious to acquire-release synchronization~\cite{Faialaa,gpuverify}). 

\noindent\textbf{\underline{Goals:}} 
Our goal is to \textit{accurately} detect data races in GPU kernels using static (compile-time) analysis. 
An accurate race detector captures true races and must minimize/eliminate false positives. 
Further, we aim to detect a broad range of races, \ie{} intra-, inter-threadblock, intra-warp, and be aware of acquire-release synchronization. 
In short, we strive to bring the best of both dynamic (\eg{} no false positives) and static (\eg{} detect infrequently manifesting races) analysis techniques. 



\mysection{A Case for Host Code Guided Race Detection} 
\label{sec:motivation:main}

We make a key observation that crucial semantic information about the values that a kernel's parameters can assume during its execution is often \emph{embedded in the host code}. 
GPU kernels do not execute in isolation -- the host (CPU) code allocates memory for the kernels, sets up the kernel's parameters, and launches the kernel with a grid of threads. 
Harnessing this hitherto ignored semantic information is critical for accurately detecting races in GPU-accelerated programs.  
We discover \textit{five classes} of information commonly found in the host code that can aid race detection for the kernels.  

\begin{figure}[t!]
\centering
\begin{code}
// Kernel code
__global__ 
void copyUppLowKernel(float *A, int rows, int cols) {
  i = blockIdx.x * blockDim.x + threadIdx.x;
  j = blockIdx.y * blockDim.y + threadIdx.y;
  if (j > i && j < rows) {
    A[j * cols + i] = A[i * cols + j];
  }
}
// Host code
void CopyUppLow(...) {
  rows = __input();  cols = __input();
  @\codebox{assert(rows == cols)}@;       // square matrices only
  copyUppLowKernel<<<#BLOCKS, #THREADS>>>(A, @\codebox{rows, cols}@);
}
\end{code}
\caption{Asserts in host code limiting kernel variable values.} 
\label{fig:host:assert}
\end{figure}

\myparagraph{\mycircled{1} Asserts in the host code}
Programmers often add assert statements to ensure that specific conditions are met in every execution of the program. 
We demonstrate how leveraging assert conditions can improve the accuracy of race detection with an example.  
\autoref{fig:host:assert} shows a simplified snippet of \textcode{\kaldia{}} program, from Kaldi~\cite{Kaldi}, a widely used library for speech recognition.
In the kernel (\textcode{copyUppLowKernel}), each GPU thread copies a value from the upper half of the matrix to the lower half (line $7$). 
This is valid only for square matrices, \ie{} \textcode{rows} and \textcode{cols} of a matrix have the same value. 
This constraint is \textit{not} encoded in the kernel but is \textit{embedded in the host code} in the form of an assert condition (line $13$). 

A host code-agnostic static analysis of the kernel code will report false positives, as is the case with all prior works. 
For example, if the values of \textcode{rows} and \textcode{cols} are $32$ and $1$, respectively, a data race could occur among threads of different threadblocks on line $7$. 
However, the assert condition in the host code ensures that this race is \textit{not} possible in any execution of the program. 
We notice similar patterns in other programs such as \textcode{\kaldib{}}~\cite{Kaldi} and \textcode{\kaldic{}}~\cite{Kaldi}. 

\begin{figure}[t!]
\centering
\begin{code}
// Kernel code
__global__ 
void reduceFinal(float *array, float *res) {
  ...
  if (threadIdx.x == 0) res[0] = ...;
}
// Host code
void reduction(float *A, ...) {
  blockReduce<<<#BLOCKS, #THREADS>>>(A, partialReduce);
  reduceFinal<<<@\codebox{1, #THREADS}@>>>(partialReduce, A); 
  // Always launched with 1 threadblock
}
\end{code}
\caption{Host code setting thread grid dimensions.} 
\label{fig:host:dimension}
\end{figure}

\myparagraph{\mycircled{2} Thread grid dimensions} 
The host code sets the thread grid dimensions during the kernel launch. 
The grid dimension put bounds on the values of thread identifiers (\eg{} \textcode{threadIdx.x}, \textcode{blockIdx.x}). 
We illustrate the usefulness of this information in race detection with the \textcode{\reduction{}} program from NVIDIA~\cite{CUDASamples} (\autoref{fig:host:dimension}) as an example. 
The host launches two kernels \textcode{blockReduce} (line $9$) and \textcode{reduceFinal} (line $10$). 
The former reduces array \textcode{A} by partitioning it into multiple chunks, which are assigned to different threadblocks (not shown). 
Each threadblock reduces the values in its partition to a single value and writes to the \textcode{partialReduce} array. 
Next, the \textcode{reduceFinal} kernel reduces the \textcode{partialReduce} array to a single value. 
One thread in the threadblock writes the final value to the $0^{th}$ index of the array (line $5$). 

Note that the host code \textit{always} launches the \textcode{reduceFinal} kernel with \textit{one} threadblock (line $10$). 
An analysis that ignores this information may assume that the kernel can be launched with multiple threadblocks. 
Threads from \textit{different} threadblocks with \textcode{threadIdx.x} as $0$ will satisfy the condition on line $5$, leading to a data race. 
However, such a race is \textit{not} possible in any execution. 
We observed a similar pattern in other programs, \eg{} \textcode{\expdist{}}~\cite{HeCBench} and \textcode{\random{}}~\cite{HeCBench}.

\begin{figure}[t!]
\centering
\begin{code}
// Kernel code
__global__ 
void toneMapping(..., int width, int channels, int output) {
  x = blockIdx.x * blockDim.x + threadIdx.x;
  y = blockIdx.y * blockDim.y + threadIdx.y;
  // compute val
  output[width * channels * y + (x * channels)] = val;
}
// Host code
void compress(..., int @\codebox{hWidth}@, int hHeight, ...) {
  // hWidth setting grid dimension and width kern. parameter
  dim3 grid(@\codebox{hWidth/16}@, hHeight/16);
  toneMapping<<<grid, ...>>>(..., @\codebox{hWidth}@, 4, ...); 
}
\end{code}
\caption{Host code creating kernel parameter relations.}
\label{fig:host:relation}
\end{figure}

\myparagraph{\mycircled{3} Relation among kernel parameters}
It is typical for the host code to derive different kernel parameters from the same variable, thereby implicitly creating a relation between the otherwise seemingly independent parameters. 
We illustrate this with the \textcode{\tone{}} program from HeCBench~\cite{HeCBench} (\autoref{fig:host:relation}).  %
It compresses a high dynamic range image to a lower range. 
In the \textcode{toneMapping} kernel, each GPU thread performs computations on RGB values of an input image (not shown) and writes the result to the output (line $7$). 
The number of threadblocks launched depends on the dimensions of the image, \ie{} variables \textcode{hWidth} and \textcode{hHeight}. 
The host computes the thread grid dimension's x-axis using \textcode{hWidth} (line $12$). 
It also sets the value of the kernel parameter \textcode{width} with the same \textcode{hWidth} variable (line $13$). 
This relation is encoded in the host code, but is \textit{not} discernible from the kernel code. 

An analysis agnostic to the relations among the kernel parameters can consider possible values of the grid dimension's x-axis and \textcode{width} as $2$ and $16$, respectively. 
These values could lead to a data race among threads of different threadblocks on line $7$. 
However, the grid dimension and \textcode{width} are computed from the same host variable (\textcode{hWidth}). 
When the grid dimension is $2$, \textcode{width} must be at least $32$ (line $12$). 
Consequently, the above perceived race is not possible. 
Many other programs, \eg{} \textcode{\convSep{}}~\cite{HeCBench}, have a similar pattern. 

\begin{figure}[t!]
\centering
\begin{code}
// Kernel code
__global__ 
void ludInternal(float *m, int dim, int offset) {
  rid = offset + (blockIdx.y + 1) * BLOCK_SIZE;
  cid = offset + (blockIdx.x + 1) * BLOCK_SIZE;
  // compute value of sum
  m[(rid + threadIdx.y) * dim  + cid + threadIdx.x] = sum;
}
// Host code
void lud(float *m, int hMatrixDim, ...) {
  // hOffset controlling values of offset kern. parameter
  @\codebox{for (hOffset = 0; hOffset < hMatrixDim; ...)}@
    ludInternal<<<...>>>(m, hMatrixDim, @\codebox{hOffset}@); 
}
\end{code}
\caption{Host code adding bounds to kernel parameters.}
\label{fig:host:control}
\end{figure}

\myparagraph{\mycircled{4} Loop bounds on parameters}
It is not  uncommon for values of kernel parameters to be bounded by loop (\eg{} \textcode{for}) iterator variables in the host code. 
Let us take the example of the \textcode{\lud{}} program (\autoref{fig:host:control}) from the Rodinia benchmark suite~\cite{rodinia}. 
It factors a matrix into the product of lower and upper triangular matrices. 
In the kernel (\textcode{ludInternal}), each thread factorizes entries of the matrix (not shown) and writes the result to the array (line $7$). 
The host iteratively launches the kernel based on the value of \textcode{hOffset}. 
It ranges from $0$ to the value of \textcode{hMatrixDim} (line $12$). 
The host also sets the value of kernel parameter \textcode{offset} with \textcode{hOffset} (line $13$). 
While this information on the bounds on the values of \textcode{offset} is \textit{not} available in the kernel, it is in the host code.

Ignoring the bounds on \textcode{offset} values (set by \textcode{hOffset}) controlled by the \textcode{for} loop in line $12$ can generate avoidable false positives.  
For example, a host-agnostic analysis may consider negative values for \textcode{offset} (\eg{} $-47$). 
Thus, a data race could occur among threads of different threadblock on line $7$. 
However, the minimum value of \textcode{offset} is $0$ thanks to the loop bound.
Thus, such a race is infeasible.
We observe similar behavior in other programs too, \eg{} \textcode{\nw{}}~\cite{HeCBench}.

\begin{figure}[t!]
\centering
\begin{code}
// kernel code
__global__
void shift(float *matrix, int rowSize, ..., texture* tex) {
  int xid = blockIdx.x * blockDim.x + threadIdx.x;
  int yid = blockIdx.y * blockDim.y + threadIdx.y;
  matrix[yid * (rowSize / 4) + xid] = tex[...];
}
// host code
int main() {
  int *matrix, hRowSize;
  cudaMallocPitch(&matrix, @\codebox{&hRowSize}@, rowBytes, numRows);
  // hRowSize is always greater than zero, i.e., positive
  shift<<<...>>>(matrix, @\codebox{hRowSize}@, ..., texture);
}
\end{code}
\caption{Allocation size in host code set as kernel parameter.}
\label{fig:host:allocation}
\end{figure}

\myparagraph{\mycircled{5} Relation between allocation size and parameters}
The host code typically sizes data structures based on the problem size (\eg{} input matrix). 
It is also common to pass the size or aspects of it (\eg{} number of columns of the matrix) as kernel parameters.
The size argument of a memory allocation request (\eg{} \textcode{cudaMalloc}, \textcode{cudaMallocPitch}) must \textit{always} be a positive value. 
If the size variable or its derivative is passed as kernel parameters, the set of values these parameters can assume in any execution is \textit{implicitly bounded}. 

We illustrate this observation with the \textcode{\texBind{}} program from NVIDIA~\cite{CUDASamples} (\autoref{fig:host:allocation}).  
In the kernel code (\textcode{shift}), each thread reads from texture memory and writes to a matrix in the global memory (line $6$). 
Threads index into the matrix using thread identifiers (\eg{} \textcode{threadIdx.x}) and the kernel parameter, \textcode{rowSize}. 
The host code allocates memory for the matrix using \textcode{cudaMallocPitch}, a specialized API for allocating matrices (line $11$). 
It requests \textcode{rowBytes} size of memory (third argument) for each of \textcode{numRows} rows (fourth argument). 
This request allocates \textcode{hRowSize} bytes (second argument) of aligned memory (\S$6.2.2$ in \cite{CUDAProgrammingGuide}) for each row. 
This argument must \textit{always} have a positive value. 
The host code then sets the value of kernel parameter \textcode{rowSize} with \textcode{hRowSize} (line $13$). 
Thus, the value of \textcode{rowSize} must also be positive. 

Ignoring such implicit semantic information can cause a static analysis technique to report false positives. 
For example, if it assumes that \textcode{rowSize} can take any possible values, including $0$, it would find a possible data race among threads from different threadblocks on line $6$. 
However, such a race \textit{cannot} occur in any execution of the program. 
We observed a similar pattern in other programs too, \eg{} \textcode{\copyRows{}}~\cite{Kaldi}. 

\noindent\textbf{\underline{Summary:}} 
We discover five classes of semantic information in the host code that affect the values that kernel parameters can assume during any execution: 
\mycircled{1} Assert conditions in the host code, 
\mycircled{2} Values of thread grid dimensions set in the host code, 
\mycircled{3} Relation among kernel parameters,
\mycircled{4} Bounds on the values due to loops in the host code, and 
\mycircled{5} Relation between allocation size and kernel parameters.
We find that leveraging these classes is imperative for accurate race detection in GPU kernels. 



\mysection{Design Principles behind \racetool{}}
\label{sec:design:main}

We design \racetool{}, a static analysis technique, to accurately detect a broad range of data races in GPU programs. 
It analyzes the kernel code to create pairs of memory access instructions that are \textit{not} separated by barrier synchronization operations (\eg{} \textcode{syncthreads}). 
It then creates constraints that must be satisfied for such instruction pairs to conflict, \ie{} access the same address from different issuing threads, and at least one of them is a write. 
Next, it adds several new constraints that limit the possible values of kernel parameters by analyzing the \textit{host} code.
These constraints must hold true in every execution of the program (\eg{} assert conditions, relations among parameters). 
\racetool{} invokes a SAT solver~\cite{z3-solver,ORToolsCPSAT} to check the satisfiability of these constraints for finding pairs of memory instructions that may race under some program input.

Next, \racetool{} checks which of these access pairs cannot race thanks to fine-grain synchronization, \ie{} acquire-release operations as in locks.  
The access pairs that remain are true data races. 
Finally, \racetool{} reports the file name, line numbers of the instructions involved in the race, and the type of race (\eg{} inter-, intra-threadblock, intra-warp) to the programmer.

We next detail our key contributions: 
\mycircled{1} Analyzing the host code to constrain possible values of kernel parameters for accurately guiding the detection of data races in the kernel code.
\mycircled{2} Enhanced analysis of the kernel code and introduction of new techniques to broaden the types of data races that a static technique can detect. 

%
\mysubsection{Host-code guided constraint generation}
\label{sec:design:host}

\racetool{} analyzes the host code to extract five classes of semantic information that could limit the possible values of kernel parameters.
It encodes them as constraints that the SAT solver must satisfy while identifying pairs of conflicting memory instructions. 
In effect, this limits the solver's search space, which allows accurate race detection. 


\myparagraph{\mycircled{1} Asserts in the host code}
\racetool{} analyzes every operation in the host code in the path from the beginning of the program, \ie{} the \textcode{main} function, till the launch of a GPU kernel.
On encountering an \textcode{assert}, it extracts the associated condition. 
It captures the variables used in the condition, the expression of how these variables are computed, and the condition (\eg{} ==) among them. 
It represents these conditions as constraints that must always be satisfied. 
For example, for the kernel in \autoref{fig:host:assert}, \racetool{} adds an equality constraint between the variables \textcode{rows} and \textcode{cols} to limit the SAT solver's search space. 

\myparagraph{\mycircled{2} Thread grid dimensions}
For every kernel launch, \racetool{} extracts the grid dimensions from the launch parameters (\eg{} each axis of \textcode{gridDim}).
It then traces how the host code computes values of these parameters. 
It adds these as constraints to the SAT solver for limiting the values of thread identifiers (\eg{} \textcode{blockIdx.x}). 
For example, to analyze the \textcode{reduceFinal} kernel in \autoref{fig:host:dimension}, \racetool{} adds the `\textcode{blockIdx.x} $<$ $1$' constraint. 
Here, \textcode{blockIdx.x} is the block identifier of a GPU thread, and the value for \textcode{gridDim.x} is set to $1$ by the host code. 
This constrains the values \textcode{blockIdx.x} can assume in the SAT solver's search space.

\myparagraph{\mycircled{3} Relation among kernel parameters}
For every variable in the host code used in computing kernel parameters, \racetool{} instantiates a solver variable. 
A solver variable represents values that a program variable can take during an execution in the SAT solver's search space. 
For every unique program variable, it instantiates a unique solver variable. 
Thus, if multiple kernel parameters depend on the same host variable, the relation among them is captured in the solver's search space.  
For example, for the kernel in \autoref{fig:host:relation}, \racetool{} creates a solver variable for \textcode{hWidth}. 
It reuses the same solver variable to create the grid dimension's x-axis and \textcode{width} parameter in the solver, \textit{preserving} their relation as the solver explores the satisfiability of the constraints.

\myparagraph{\mycircled{4} Loop bounds on parameters}
\racetool{} traces all host variables that any of the kernel parameters depend on. 
Suppose there is a loop iterator among them. 
It then adds constraints to limit the values of the corresponding kernel parameter by the loop bounds. 
For example, for the kernel in \autoref{fig:host:control}, \racetool{} adds `\textcode{offset} $\geq$ $0$' and `\textcode{offset} $<$ \textcode{hMatrixDim}' constraints to the SAT solver. 
These constraints bound the values that \textcode{offset} can assume in the SAT solver's search space. 

\myparagraph{\mycircled{5} Allocation size and parameters}
\racetool{} identifies the host variables used in both sizing memory allocations and passed as kernel parameters. 
For each memory allocation (\eg{} \textcode{cudaMalloc}, \textcode{cudaMallocPitch}), it extracts the host variables that determine the size of the allocation. 
If the same variable is passed as a kernel parameter, it constrains the possible values of the parameter to be positive. 
For example, for the kernel in \autoref{fig:host:allocation}, \racetool{} adds `\textcode{rowSize} $> 0$' constraint. 
This ensures that \textcode{rowSize} is positive in the solver's search space. 

\mysubsection{Broadening the ambit of race detection}
\label{sec:design:kernel}

\racetool{} analyzes the kernel code and adds constraints to detect the presence (or absence) of intra-warp races. 
Further, it adds constraints and invokes a SAT solver to find the conflicting memory accesses that are adequately guarded by acquire-release synchronization and those that are \textit{not} (true races). 

\myparagraph{Intra-warp races}
\racetool{} checks if memory instructions in a pair are separated by a warp barrier instruction (\textcode{syncwarp}) in all program paths.
Specifically, it checks if: 
\mycircled{1} A barrier post-dominates~\cite{Aho2006Compilers,dom1,dom2} the first instruction, and 
\mycircled{2} The barrier dominates the second instruction. 
If true, the pair can never participate in an \textit{intra}-warp race. 
Otherwise, it adds constraints to the SAT solver to check if the pair of instructions can be issued by different threads belonging to the same warp.
If yes, it declares an intra-warp race. 


\myparagraph{Acquire-release synchronization}
For every pair of conflicting memory access instructions identified by the SAT solver, \racetool{} checks if acquire-release synchronization operations avoid a race between the instructions.    
Specifically, it checks if the conflicting accesses are encompassed within the lock acquire and lock release operations. 
If not, it further checks if the accesses are part of producer-consumer communication.

For each instruction (data access) in the conflicting pair, \racetool{} back-traces the kernel code to find the acquire that dominates the instruction (if it exists). 
Similarly, it forward-traces to find the release operation that post-dominates it. 

If \racetool{} fails to find an encompassing lock acquire and release, it checks for the producer-consumer-based pattern. 
Here, the conflicting pair consists of a producer of data (write instruction) and a consumer of data (read instruction).
However, they cannot race if a release operation follows the producer write while a matching acquire precedes the consumer.

The existence of an encompassing lock acquire and release or producer-consumer pattern, by itself, does \textit{not} guarantee the absence of race among the conflicting pair of memory instructions.
A race can still exist if the address of the acquire and release fail to match (\eg{} different lock addresses). 
\racetool{} relies on the SAT solver to determine if addresses of the acquire and release can mismatch under any possible program input. 
If yes, it declares a race. 
Otherwise, not.

\mysection{Implementation of \racetool{}}
\label{sec:impl:main}

\begin{figure}[t!]
    \centering
    \includegraphics[width=\linewidth]{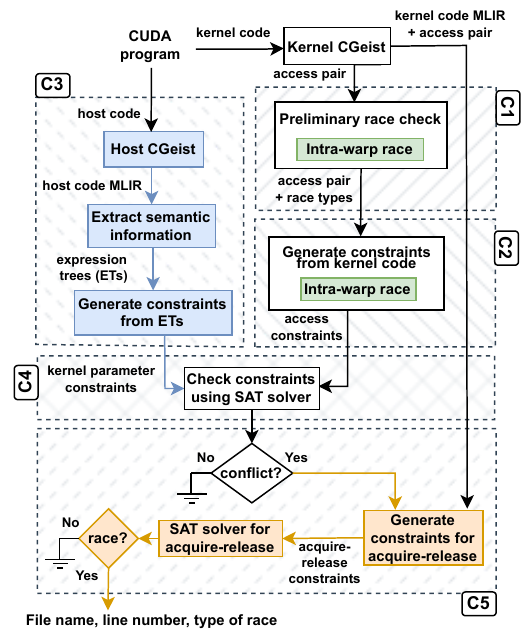}
    \caption{High-level components of \racetool{}. Colored (blue, green and orange) elements are introduced by \racetool{}.}
    \label{fig:impl:main}
\end{figure}

\autoref{fig:impl:main} depicts the high-level working of \racetool{}, highlighting our key contributions with \emph{colored elements}. 
It compiles a \cuda{} program consisting of the host and GPU kernel code using CGeist~\cite{Polygeist} to MLIR intermediate representation~\cite{MLIR}.  
MLIR enables easy analysis of the code.

\racetool{} has five major components: 
\component{1} It analyzes kernel code's MLIR to find pairs of memory access instructions that can never race (\eg{} pairs of loads or atomics).
\component{2} For the pairs that pass this preliminary check, \racetool{} generates constraints that must be satisfied for its constituent instructions to race, \ie{} they may access the same address and can be issued by different threads.
\component{3} Uniquely, \racetool{} also analyzes host code to generate additional constraints that must be satisfied in all program executions (\eg{} relation among kernel parameters). 
\component{4} It uses Google OR-Tools' CP-SAT solver~\cite{ORToolsCPSAT} to find if these constraints are satisfiable, thereby identifying conflicting access pairs that \textit{may} race. 
\component{5} Finally, \racetool{} checks if the kernel contains acquire-release synchronization and if that prevents potential races. 
Otherwise, it reports every racey memory access pair with the corresponding filename, line number, and the type of race to the programmer. 

In this section, we emphasize \racetool{}'s key contributions, as described in \autoref{sec:design:main}.
\mycircled{1} Host-code guided constraint generation is in \component{3}.
\mycircled{2} The enhancement for detecting intra-warp races is spread between \component{1} and \component{2}.
The support for race detection in the presence of acquire-release synchronization is encapsulated in \component{5}.

\racetool{} also supports scoped-synchronization operations~\cite{ScoRD,iGUARD}. 
For brevity, we describe its implementation considering the default scope (device) for those operations. 
We discuss the support of scopes in the Appendix. 

\mysubsection{Preliminary race check}
\label{sec:impl:preliminary}


Even in programs with races, most pairs of memory access instructions do not conflict and thus, cannot race~\cite{ScoRD,iGUARD}. 
For efficiency, \racetool{} performs preliminary checks to ascertain if an instruction pair can conflict or if they are \textit{trivially} race-free~\cite{iGUARD} (\component{1} in \autoref{fig:impl:main}).

\myparagraph{Generating access pairs}
We denote every static memory access instruction (loads, stores, and atomics) as an \access{}.
\racetool{} generates a list of \accesspair{}s from all memory instructions in the kernel code that access the same data structure using Kernel CGeist (the top of \autoref{fig:impl:main}). 

\myparagraph{Avoiding race checks}
Races occur when \access{}es issued by different threads alias to the same address, at least one is write, and there is no intervening synchronization. 
\racetool{} examines three conditions to decide if an \accesspair{} can participate in any of the three types of race or is trivially race-free. 

\begin{mybullet}
%
\item \textbf{Avoiding intra-warp check.}
\racetool{} introduces this condition (green in \component{1}, \autoref{fig:impl:main}). 
An \accesspair{} cannot participate in an intra-warp race if a warp barrier (\textcode{syncwarp}) is present in \textit{all} program paths between them. 
Suppose the \accesspair{} consists of instructions A\textsubscript{i} and A\textsubscript{j}  (A\textsubscript{j} follows A\textsubscript{i} in program order).
\racetool{} checks if a warp barrier in kernel code's MLIR post-dominates A\textsubscript{i} (\ie{} all program paths from Ai to the end of kernel code) \emph{and} dominates A\textsubscript{j} (\ie{} all program paths from the beginning of the kernel code to A\textsubscript{j}). 
We use MLIR's standard routines to check the dominance relations~\cite{mlirDomInfo}. 
If such a barrier is present, \racetool{} performs inter- and intra-threadblock race checks for the pair. 
%
\item \textbf{Avoiding intra-threadblock and intra-warp checks.}
An \accesspair{} cannot participate in an intra-threadblock and intra-warp race if a block barrier (\textcode{syncthreads}) is present in \textit{all} program paths between the constituent \access{}es. 
A block barrier synchronizes all the threads of a threadblock, which includes threads of a warp. 
Thus, it prevents intra-warp races too. 
For this, \racetool{} uses a method similar to the one described earlier, \ie{} dominance relation between \accesspair{} and a block barrier. 
If such a barrier is present, \racetool{} performs only inter-threadblock race check for the pair. 
%
\item \textbf{Trivially race-free.} 
An \accesspair{} can never race if the constituent \access{}es are both either load or atomic instructions. 
\racetool{} performs no further race checks for the pair. 
\end{mybullet}

After this step, \racetool{} is left with a subset of pairs that \textit{may} conflict, and types of race(s) they may participate in. 

\begin{table}[t!]
    \footnotesize
    \centering
    \caption{Methods \racetool{} uses to detect races. \high{Highlighted} methods are introduced by \racetool{}.}
    \label{tab:impl:methods}
    \vspace{-.5em}
    \begin{tabular}{c|l}\hline
   \textbf{Method}                & \textbf{Expectation}                                       \\ \hline
   addr(A)                        & possible addresses for \access{} A                         \\ \hline
   tb(A)                          & threadblock identifier of a thread issuing \access{} A     \\ \hline
   id(A)                          & thread identifier of a thread within its threadblock       \\ 
                                  & issuing \access{} A                                        \\ \hline
   \high{wp(A)}                   & warp identifier of a thread issuing \access{} A             \\ \hline
   \high{syn(A)}                  & acquire/release \access{} for \access{} A                  \\ \hline
    \end{tabular}
\end{table}

\mysubsection{Generate constraints from kernel code}
\label{sec:impl:constraint}

\racetool{} analyzes each \accesspair{} that passes preliminary race checks to determine if it can conflict (\component{2} in \autoref{fig:impl:main}). 
Specifically, it creates two sets of constraints for each \accesspair{}. 
The first set checks the possibility of addresses of the instructions in the pair to alias, \ie{} \emph{addr(A\textsubscript{i}) == addr(A\textsubscript{j})} (notations in \autoref{tab:impl:methods}). 
The second set checks if different threads can issue the constituent instructions.

If preliminary checks determined that the given pair can potentially participate in an intra-warp race, \racetool{} adds a constraint to check if the pair can be issued by different threads belonging to the same warp ($wp(A\textsubscript{i})$ == $wp(A\textsubscript{j})$). 
This constraint is newly introduced by \racetool{} (green in \component{2}, \autoref{fig:impl:main}). 
\racetool{} creates solver variables that represent the warp identifiers of the \accesspair{} (say, \textit{solver}WpA\textsubscript{i} and \textit{solver}WpA\textsubscript{j}).
It adds an equality constraint between them, \ie{} \textit{solver}WpA\textsubscript{i} == \textit{solver}WpA\textsubscript{j}. 
Further, the threads issuing the pair must belong to the same threadblock ($id(A\textsubscript{i})$ $\neq$ $id(A\textsubscript{j})$, $tb(A\textsubscript{i})$ == $tb(A\textsubscript{j})$). 
\racetool{} also adds constraints to ensure this condition. 

Similarly, suppose the preliminary checks determined that the given pair can potentially participate in an inter-threadblock race. 
\racetool{} adds a constraint to check if the instructions can be issued by threads from different threadblocks ($tb(A\textsubscript{i}) \neq tb(A\textsubscript{j})$).
If intra-threadblock race is possible, the constraints check if the pair of instructions can be issued by different threads belonging to the same threadblock ($id(A\textsubscript{i}) \neq id(A\textsubscript{j})$, $tb(A\textsubscript{i})$ == $tb(A\textsubscript{j})$, $wp(A\textsubscript{i}) \neq wp(A\textsubscript{j})$). 
These constraints are then provided to the SAT solver (\component{4}). 

\mysubsection{Generate constraints from host code}
\label{sec:impl:analysis:host}
\racetool{} analyzes the host code's MLIR to generate new constraints that limit the range of values that kernel parameters can assume (\component{3} in \autoref{fig:impl:main}) -- a first-of-its-kind feature in this space. 
It does so in two steps. 
First, it generates expression trees (ET)~\cite{expressionTrees} that involve host variables affecting the values of kernel parameters. 
Next, it converts the ETs to constraints that must simultaneously hold true in every execution of the program and that the SAT solver must satisfy while finding conflicting \accesspair{}s.

\begin{mybullet}
\item \textit{Extract semantic information} in the form of ET. 
An ET captures the expressions (equations) that compute an operation (\eg{} conditions) or the values a variable can assume during execution. 
ETs are organized as binary trees, where the leaf nodes represent variables (\eg{} constants, user inputs) and internal nodes consist of binary operations (\eg{} `$+$') on their children. 
The root of the tree captures how an operation is computed or a variable is initialized. 
Nodes in ET also track the MLIR \textcode{Value} objects~\cite{mlirValRef} -- a unique identifier for variables in MLIR representation of the code. 

\begin{figure}[t!]
    \centering
    \includegraphics[width=.4\linewidth]{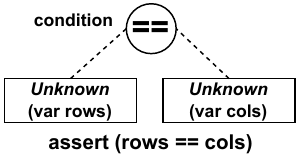}
\caption{Expression tree for assert condition in \autoref{fig:host:assert}.}
\label{fig:impl:et}
\end{figure}

As an example, \autoref{fig:impl:et} shows the ET for the assert condition in \autoref{fig:host:assert}, line $13$. 
Here, \textcode{rows} and \textcode{cols} must be equal. 
The leaf nodes capture information about \textcode{rows} and \textcode{cols}, \ie{} they have \textit{Unknown} value as they are user inputs. 
The root node captures the condition among the two leaf nodes (==). 

\begin{figure}[t!]
\begin{subfigure}{.5\textwidth}
\centering
\begin{code}
ET createET(Op operation) {
  // recursively create ET for operation and its constituents
  if (isConstantOperation(operation)) {        // constant ET
    return ET { operation.ConstantValue };
  } else if (isUserInput(operation)) {          // unknown ET
    return ET { Unknown };
  } else if (isBinaryOperation(operation)) {     // binary ET
    leftET = createET(operation.lhs);
    rightET = createET(operation.rhs);
    return ET { leftET operation.op rightET }; 
  } else if (isForOperation(operation)) {        // bounds ET
    minET = createET(operation.getLowerBound());
    maxET = createET(operation.getUpperBound());
    return ET { minET maxET };
  } else if (/* other possibilities */) { ... }
}
\end{code}
\caption{Simplified code snippet to create expression trees.}
\vspace{.5em}  
\label{fig:impl:express:create}
\end{subfigure}
\hfill
\begin{subfigure}{.5\textwidth}
\centering
\begin{code}
vector<ET> asserts;                      // tracking asserts
// tracking a kernel's grid dimension and arguments
tuple<Kernel, vector<ET>, vector<ET>> kernelInfo;
vector<Value> allocVars;                 // allocations
if (operation == Assert) {               // assert conditions
  asserts.push_back(createET(operation.condition()));
} else if (operation == KernelLaunch) {
  for (dim in operation.GridDim())       // kernel dimensions
    gridDims.push_back(createET(dim));
  for (arg in operation.KernelArgs())    // kernel arguments
    args.push_back(createET(arg));
  kernelInfo = { operation.KernelName, gridDims, args };
} else if (operation == AllocOp) {       // memory allocation
  sizeVar = backTrace(operation.arg[sizeArg]);
  allocVars.push_back(sizeVar.Value);
} else if (/* other host code operations */) { ... }
\end{code}
\caption{Simplified code snippet for host code analysis.}
\label{fig:impl:express:host}
\end{subfigure}
\caption{Creating expression trees from host code analysis.}
\vspace{-1em}
\label{fig:impl:express}
\end{figure}

\autoref{fig:impl:express:create} shows the simplified code snippet that creates ETs (\textcode{createET}) for an MLIR operation (\eg{} a condition). 
It recursively creates ETs for its constituents, \eg{} the left and right operands of a binary operation (lines $8$-$9$), until it encounters a constant or \textit{Unknown} (\ie{} user input). 
\autoref{fig:impl:express:host} shows the simplified code snippet that analyzes the host code's MLIR to capture ETs by invoking \textcode{createET} upon encountering operations of interest (\eg{} asserts, kernel launch, memory allocations). 
\item \textit{Generate constraints from ETs} for the SAT solver. 
\racetool{} expresses constraints in the form of equalities or inequalities among \textit{solver variables} and/or constants.
The solver variables represent program variables in the SAT solver's search space. 
\end{mybullet}

We now elaborate these two steps for each class of semantic information extracted from the host code (\autoref{sec:motivation:main}). 

\myparagraph{\mycircled{1} Asserts in host code}
For every assert in the host code, \racetool{} creates an ET for the assert condition. 
We describe this step by tracing \autoref{fig:impl:express:create} with the \textcode{assert} from \autoref{fig:host:assert}, line $13$. 
Asserts are typically binary relations (\eg{} ==). 
\racetool{} recursively creates ETs for the left and right operands of the binary relation of the assert's condition (lines $7$-$10$). 
The left operand \textcode{rows} is a user input. 
Thus, it returns an ET of \textit{Unknown} type, \ie{} any value (line $6$).  
Similarly, the right operand \textcode{cols} is an \textit{Unknown} type. 
Finally, \racetool{} returns an ET for the condition, which can be visualized as `\textit{Unknown}Rows == \textit{Unknown}Cols.'  
While the actual values of \textcode{rows} and \textcode{cols} are not known during compilation, their relation (equality) is the key information that \racetool{} harnesses. 
It then adds this ET to the list of asserts (\autoref{fig:impl:express:host}, line $6$). 

To generate constraints for the SAT solver from the ET, \racetool{} creates solver variables for the corresponding program variables in the assert condition, \ie{} \textit{solver}Rows and \textit{solver}Cols here. 
To preserve the condition, it creates an equality constraint between the solver variables, \ie{} \textit{solver}Rows == \textit{solver}Cols, that the solver must satisfy. 

\myparagraph{\mycircled{2} Thread grid dimension set by host}
When \racetool{} encounters a kernel launch (\textcode{LaunchOp} in MLIR), it creates ETs for each axis of the grid and block dimensions. 
For explanation, we trace the snippet from \autoref{fig:impl:express:create} with the grid dimensions for the kernel launch in \autoref{fig:host:dimension}, line $10$ (block dimensions are captured similarly). 
A kernel's grid dimensions are typically integers (\eg{} constants) or binary arithmetic operations (\eg{} $+$). 
Here, the x-axis is a constant with value $1$, satisfying the condition on line $3$. 
\racetool{} returns an ET which can be visualized as `\textit{Constant} with value $1$.' 
It similarly creates an ET for the y and z axes (\autoref{fig:impl:express:host}, lines $8$-$9$). 

\racetool{} uses grid dimensions to constrain the values of thread identifiers that each thread in a kernel can assume during execution. 
For example, the threadblock identifier (\ie{} \textcode{blockIdx.x}) of each thread must be less than the grid dimension. 
Thus, it creates three solver variables, one for \textcode{gridDim.x} ($1$, constant here), and one for each \access{} (\textit{solver}BlockA1 and \textit{solver}BlockA2).
It then generates less than constraints between the corresponding solver variables, \eg{} \textit{solver}BlockA1 $<1$, that the SAT solver must satisfy. 

\myparagraph{\mycircled{3} Relation among kernel parameters}
While creating an ET for a kernel parameter, \racetool{} captures all the variables in the host code used in computing the parameter through their unique MLIR \textcode{Value} object. 
For example, for \autoref{fig:host:relation}, \racetool{} tracks the variable \textcode{hWidth} (line $10$) with its \textcode{Value} object (say, \textit{Val}HWidth). 
While creating ETs for \textcode{width} kernel parameter and grid dimension's x-axis, it observes that they both use \textcode{hWidth}. 
\racetool{} captures this information in both the ETs. 
The ETs for \textcode{width} and grid dimension's x-axis can be visualized as `\textit{Val}HWidth' and `\textit{Val}HWidth / 16,' respectively. 

While creating a solver variable for a kernel parameter, \racetool{} creates solver variables for each variable in the parameter's ET. 
It maintains a \emph{map} for each variable to its corresponding solver variable. 
All parameters share this map. 
If a variable is absent from the map (\ie{} first use), it creates a corresponding solver variable and adds it to the map. 
Otherwise, it \textit{reuses} the corresponding solver variable. 
This reflects the relation among the parameters on the corresponding solver variables. 
For example, the ETs of \textcode{width} and grid dimension's x-axis in \autoref{fig:host:relation} have \textcode{hWidth} in them. 
While creating a solver variable for \textcode{width}, the variable \textcode{hWidth} is absent from the map.
\racetool{} then creates a solver variable for \textcode{hWidth} and adds it to the map. 
While creating a solver variable from the ET of the grid dimension's x-axis, it finds \textcode{hWidth} in the map and reuses it, preserving the relation among parameters.  

\myparagraph{\mycircled{4} Loop bounds on parameters}
While creating an ET for a kernel parameter, \racetool{} checks if it depends on the iterator variable of a loop (\eg{} \textcode{forOp} in MLIR). 
If yes, it creates an ET for the iterator's bounds, \ie{} minimum and maximum values that the iterator can assume.
It associates these bounds with the kernel parameter. 
For explanation, we trace the snippet from \autoref{fig:impl:express:create} with the \textcode{offset} kernel parameter from \autoref{fig:host:control}. 
\racetool{} observes that \textcode{offset} is the iterator variable (\textcode{hOffset}) of a \textcode{for} loop, satisfying the condition on line $11$. 
It creates an ET for the minimum and maximum values of \textcode{hOffset}, and associates them with \textcode{offset}. 
The ET for \textcode{offset} can be visualized as `minET maxET' where minET and maxET are `\textit{Constant} with value $0$,' and `\textit{Unknown}HMatrixDim' respectively (\textcode{hMatrixDim} is a user input, thus, unknown). 
\racetool{} captures this ET as part of kernel arguments (\autoref{fig:impl:express:host}, lines $10$-$11$). 

\racetool{} constrains the value that \textcode{offset} can assume due to loop bounds. 
It creates three solver variables, one for the kernel parameter (\textit{solver}Offset), and one each for minimum ($0$, constant here) and maximum values (\textit{solver}HMatDim) of the loop. 
Here, \textit{solver}HMatDim is the solver variable associated with the variable \textcode{hMatrixDim}. 
It then generates constraints that bound the values of the kernel parameter, \ie{} \textit{solver}Offset $\geq$ $0$ and \textit{solver}Offset $<$ \textit{solver}HMatDim.

\myparagraph{\mycircled{5} Relation between allocation size and parameters}
For every memory allocation operation in the host code's MLIR (\eg{} \textcode{cudaMallocPitch}), \racetool{} captures the variable that contains the size of the allocation. 
We describe this procedure by tracing the snippet in \autoref{fig:impl:express:host} with the memory allocation on line $11$ in \autoref{fig:host:allocation}. 
The allocation satisfies the condition on line $13$. 
\racetool{} then back traces the size argument to find the variable involved in computing the argument. 
In this case, the variable is \textcode{hRowSize} on line $14$. 
It captures this variable by adding the corresponding MLIR \textcode{Value} object to the \textcode{allocVars} list (line $15$). 

While creating solver variables for kernel parameters, \racetool{} checks if it uses a variable in the \textcode{allocVars} list. 
For this, it compares the variables (\ie{} \textcode{Value}) in the parameter's ET with each \textcode{Value} in the \textcode{allocVars} list. 
If present, \racetool{} adds a constraint on the corresponding solver variable that it must be greater than zero (positive). 
For example, in \autoref{fig:host:allocation}, the \textcode{rowSize} kernel parameter is set using \textcode{hRowSize}, which is present in the \textcode{allocVars} list. 
\racetool{} then creates a solver variable for \textcode{rowSize} and adds a greater than zero constraint on it, \ie{} \textit{solver}RowSize $>0$. 

All additional constraints generated by analyzing the host code's MLIR are provided to the SAT solver (\component{4}). 

\mysubsection{Check constraints using SAT solver}
\label{sec:impl:sat}

\racetool{} invokes the SAT solver (\component{4} in \autoref{fig:impl:main}) to check the satisfiability of the the constraints provided from kernel code analysis (\component{2}) and host code analysis (\component{3}). 
It performs this for each \accesspair{} that passed preliminary checks. 

Any \accesspair{} whose constraints are found satisfiable is declared as a conflicting pair of memory instructions that may participate in races. 
Otherwise, the \accesspair{} cannot race.


\mysubsection{Support acquire-release synchronization} 
\label{sec:impl:lock}

Finally, \racetool{} checks if the conflicting \accesspair{} identified by the SAT solver (\component{4}) can truly race (\component{5} in \autoref{fig:impl:main}). 
Note that the constraints discussed so far are agnostic to acquire-release synchronization (\eg{} locks). 
Thus, the \accesspair{} may not race if it is adequately guarded by acquire-release. 

\racetool{} checks if the conflicting \accesspair{} truly races in two steps: 
\mycircled{1} For each \access{} in a pair, it finds corresponding acquire-release operations in the kernel (if they exist). 
\mycircled{2} It creates constraints for the SAT solver to check if the address of the acquire/release may \textit{mismatch}. 

\begin{figure}[t!]
\centering
\begin{pycode}
def syn(access): # return acquire/release for access
   # STEP1. check for lock-based synchronization
   acquire = getAcquire(access)
   release = getRelease(access)
   # acquire, release must map to same address
   if match(acquire, release):
      return acquire   # or release for lock
   # STEP2. check for producer consumer pattern
   if access is WRITE and release is found: # from line 4
      return release   # release for producer
   if access is READ and acquire is found:  # from line 3
      return acquire   # acquire for consumer
   return NULL               # not guarded by acquire-release
\end{pycode}
\vspace{-1.25em}
\caption{Getting acquire/release for an \access{}.}
\vspace{-.25em}
\label{fig:impl:lock}
\end{figure}

\begin{table}[t!]
    \scriptsize
    \centering
    \caption{Constraints to detect races among memory \access{}es A\textsubscript{i}, Aj, using methods from \autoref{tab:impl:methods}. \high{Highlighted} constraints are the ones introduced by \racetool{}.}
    \label{tab:impl:rules}
    \begin{tabular}{c|c}\hline
       \textbf{Type of race}         & \textbf{Constraints}                                                                                                   \\ \hline
       Inter-threadblock             & addr(A\textsubscript{i}) == addr(A\textsubscript{j}), tb(A\textsubscript{i}) $\neq$ tb(A\textsubscript{j}),            \\
                                     & \high{addr(syn(A\textsubscript{i})) $\neq$ addr(syn(A\textsubscript{j}))}                                              \\ \hline
       Intra-threadblock             & addr(A\textsubscript{i}) == addr(A\textsubscript{j}), tb(A\textsubscript{i}) == tb(A\textsubscript{j}),                \\
                                     & id(A\textsubscript{i}) $\neq$ id(A\textsubscript{j}), wp(A\textsubscript{i}) $\neq$ wp(A\textsubscript{j}),            \\
                                     & \high{addr(syn(A\textsubscript{i})) $\neq$ addr(syn(A\textsubscript{j}))}                                              \\ \hline
       Intra-warp                    & \high{addr(A\textsubscript{i}) == addr(A\textsubscript{j}), tb(A\textsubscript{i}) == tb(A\textsubscript{j}),}         \\
                                     & \high{id(A\textsubscript{i}) $\neq$ id(A\textsubscript{j}), wp(A\textsubscript{i}) == wp(A\textsubscript{j}),}         \\
                                     & \high{addr(syn(A\textsubscript{i})) $\neq$ addr(syn(A\textsubscript{j}))}                                              \\ \hline
    \end{tabular}
\end{table}

\myparagraph{Lock/unlock} For each (data) \access{} in a conflicting pair, \racetool{} invokes \textit{syn} (\autoref{fig:impl:lock}, listed in \autoref{tab:impl:methods}) to find the corresponding acquire/release synchronization operation (\textcode{NULL} if not present). 
First, it checks if the \access{} is encompassed within lock acquire and lock release operations. 
It invokes \textcode{getAcquire} (line $3$) to find an acquire that dominates the data \access{}. 
In \cuda{}, an acquire is formed by an \textcode{atomicCAS} instruction followed by a \textcode{fence}~\cite{CUDAProgrammingGuide}. 
The method (not shown) back-traces from the \access{} to find a \textcode{fence} and then an \textcode{atomicCAS} that both dominate the \access{}. 
Similarly, \textcode{getRelease} (line $4$) looks for a release that post-dominates the \access{}. 
In \cuda{}, a release is a \textcode{fence} followed by an \textcode{atomicExch}~\cite{CUDAProgrammingGuide}. 
The method (not shown) forward-traces kernel code MLIR from the \access{} to find a \textcode{fence} and then an \textcode{atomicExch} that post-dominate the \access{}. 
\racetool{} then checks whether the acquire and release match (line $6$), \ie{} they have the same address. 
On a match, it returns the identity of the acquire operation (line $7$). 

\myparagraph{Producer/consumer} If \access{}es are not encompassed by lock acquire and release, \racetool{} checks if they participate in a synchronized producer-consumer pattern.   
Here, the conflicting \accesspair{} consists of write (producer) and read (consumer) \access{}es. 
For the write access, if a release \access{} is found (line $9$), it returns the release (line $10$). 
For the read \access{}, if the acquire \access{} is found (line $11$), it returns the acquire (line $12$). 
Otherwise, the \textcode{syn} method returns a \textcode{NULL}. 
This return value indicates that the \access{} is not guarded by acquire-release (line $13$). 

\myparagraph{Matching synchronization address} Existence of acquire-release does not guarantee freedom from races.
The address of acquire/release of the conflicting accesses must match (\eg{} same lock address).
Otherwise, the pair can still race. 
\racetool{} checks this by invoking the SAT solver again. 

Let S\textsubscript{i} and S\textsubscript{j} be acquire/release returned by the \textcode{syn} method for the \access{}es A\textsubscript{i} and A\textsubscript{j} of a conflicting \accesspair{}. 
If either of S\textsubscript{i} or S\textsubscript{j} is \textcode{NULL}, the \accesspair{} can participate in a race. 
Otherwise, it uses the SAT solver to ascertain if the addresses of S\textsubscript{i} and S\textsubscript{j} may \textit{not} match (alias) under some input. 
Specifically, \racetool{} creates solver variables from the indices of S\textsubscript{i} (\textit{solver}Sync\textsubscript{i}) and S\textsubscript{j} (\textit{solver}Sync\textsubscript{j}) that represent the addresses they can assume. 
It then creates an inequality constraint between them, \ie{} \textit{solver}Sync\textsubscript{i} $\neq$ \textit{solver}Sync\textsubscript{j}. 

If the SAT solver finds this constraint feasible, \racetool{} declares the \accesspair{} to be racey. Otherwise, not. 
\autoref{tab:impl:rules} lists all the constraints for detecting each type of data race. 

\mysection{Evaluation}
\label{sec:eval:main}

Our evaluation has the following objectives: 
\mycircled{1} Understand \racetool{}'s ability to accurately detect data races in GPU programs and quantitatively compare it with closely related works,
\mycircled{2} Analyze the usefulness of host code-guided analysis in improving the accuracy of race detection, and
\mycircled{3} Understand the overheads of \racetool{}. 

Our evaluation uses programs from popular open-source repositories such as Kaldi~\cite{Kaldi}, HeCBench~\cite{HeCBench}, NVIDIA samples~\cite{CUDASamples}, and Indigo~\cite{Indigo}, Rodinia~\cite{rodinia}, ScoR~\cite{scor} benchmark suites. 
They cover a wide range of memory access (\eg{} iterative, graph-based, random) and synchronization patterns (\eg{} acquire-release, atomics, and barriers). 
We quantitatively compare against \iguard{}~\cite{iGUARD}, a dynamic tool to detect races, and static analysis tools \gverify{}~\cite{gpuverify}, \faial{}~\cite{Faialaa}.
We also compare against \gklee{}~\cite{GKLEE}~\cite{gkle-test}, a dynamic tool that selectively employs symbolic execution (static). 

Our experimental platform has an AMD Ryzen 9 CPU with $128$GB DRAM and an NVIDIA RTX 3090 GPU.
The software stack consisted of Ubuntu $20.04$ OS, CUDA version $11.2$, and NVIDIA driver $470$.
We will open-source \racetool{}.

\mysubsection{Analyzing GPU race detection accuracy}
\label{sec:eval:analysis}


\begin{table}
\centering
\scriptsize
\caption{Race detection results of multiple tools (NS - not supported). Each cell mentions races reported with false negatives, false positives in brackets (separated by ``:'').}
\label{tab:eval:main}
\setlength{\tabcolsep}{3pt}
\begin{tabular}{l|c|c|c|c|c} \hline
    \textbf{Program Name}                    & \textbf{\iguard{}}     & \textbf{\gklee{}} & \textbf{\gverify{}} & \textbf{\faial{}} & \textbf{\racetool{}}               \\ \hline
    \indigoa{}~\cite{Indigo}                    &     0 [2:0]            &     2 [0:0]       &     2 [0:0]      &     1 [0:0]   &     2 [0:0]            \\
    \indigoc{}~\cite{Indigo}                    &     0 [2:0]            &     0 [2:0]       &     1 [0:0]      &     1 [0:0]   &     2 [0:0]            \\
    \indigob{}~\cite{Indigo}                    &     0 [2:0]            &     0 [2:0]       &     1 [0:0]      &     1 [0:0]   &     2 [0:0]            \\
    conel~\cite{GKLEE}                          &     2 [0:0]            &     2 [0:0]       &     1 [0:0]      &     1 [0:0]   &     2 [0:0]            \\
    warpCommunicate~\cite{gkle-test}            &     2 [0:0]            &     NS            &     NS           &     0 [2:0]   &     2 [0:0]            \\
    tissue~\cite{HeCBench}                      &     1 [3:0]            &     NS            &     3 [0:0]      &     0 [4:0]   &     4 [0:0]            \\
    1dconv~\cite{scor}                          &     1 [0:0]            &     2 [0:1]       &     1 [0:0]      &     1 [0:0]   &     1 [0:0]            \\ \hline
    bilateralFilter~\cite{HeCBench}             &     0 [0:0]            &     NS            &     1 [0:1]      &     0 [0:0]   &     0 [0:0]            \\
    \kaldia{}~\cite{Kaldi}                      &     0 [0:0]            &     0 [0:0]       &     2 [0:2]      &     1 [0:1]   &     0 [0:0]            \\
    \kaldib{}~\cite{Kaldi}                      &     0 [0:0]            &     0 [0:0]       &     2 [0:2]      &     1 [0:1]   &     0 [0:0]            \\
    \kaldic{}~\cite{Kaldi}                      &     0 [0:0]            &     1 [0:1]       &     2 [0:2]      &     1 [0:1]   &     0 [0:0]            \\
    \reduction{}~\cite{CUDASamples}             &     0 [0:0]            &     1 [0:1]       &     1 [0:1]      &     1 [0:1]   &     0 [0:0]            \\ 
    \random{}~\cite{RandomAccess}               &     0 [0:0]            &     NS            &     4 [0:4]      &     1 [0:1]   &     0 [0:0]            \\
    \expdist{}~\cite{HeCBench}                  &     0 [0:0]            &     1 [0:1]       &     1 [0:1]      &     1 [0:1]   &     0 [0:0]            \\
    \tone{}~\cite{HeCBench}                     &     0 [0:0]            &     0 [0:0]       &     4 [0:4]      &     1 [0:1]   &     0 [0:0]            \\
    \convSep{}~\cite{CUDASamples}               &     0 [0:0]            &     0 [0:0]       &     1 [0:1]      &     1 [0:1]   &     0 [0:0]            \\
    \lud{}~\cite{rodinia}                       &     0 [0:0]            &     NS            &     2 [0:2]      &     1 [0:1]   &     0 [0:0]            \\
    \nw{}~\cite{HeCBench}                       &     0 [0:0]            &     NS            &     2 [0:2]      &     2 [0:2]   &     0 [0:0]            \\
    \texBind{}~\cite{CUDASamples}               &     0 [0:0]            &     0 [0:0]       &     2 [0:2]      &     2 [0:2]   &     0 [0:0]            \\ 
    \copyRows{}~\cite{Kaldi}                    &     0 [0:0]            &     0 [0:0]       &     1 [0:1]      &     1 [0:1]   &     0 [0:0]            \\ 
    matMul~\cite{scor}                          &     0 [0:0]            &     NS            &     5 [0:5]      &     1 [0:1]   &     0 [0:0]            \\ 
    rule110~\cite{scor}                         &     0 [0:0]            &     NS            &     5 [0:5]      &     1 [0:1]   &     0 [0:0]            \\ \hline
    \textbf{Summary}                            &  \textbf{6 [9:0]}      &  \textbf{9 [4:4]} & \textbf{44 [0:35]}  &    \textbf{21 [6:16]}   &   \textbf{15 [0:0]}      \\ \hline
\end{tabular}
\end{table}

\autoref{tab:eval:main} presents the number of races reported on a diverse set of programs by different tools --  \iguard{}, \gklee{}, \gverify{}, \faial{}, and \racetool{}.
The top part of the table contains (racey) programs containing true data races, while the second part contains programs with no race.
The racey programs help evaluate a tool's ability to detect true races, while the ones without races help understand the propensity of different tools to report false positives.
Each cell of the table also reports the number of false negatives (say, ``x'') and false positives (``y'') reported by a given tool for a given program in the ``[x:y]'' format. 
The last row of the table summarizes the total number of races reported by a given tool along with its accuracy, \ie{} the total number of false negatives and positives.
A lower number of both false negatives and false positives indicates better accuracy in race detection and is desirable.  

\iguard{}, a dynamic analysis tool, misses true races in four programs, \ie{} false negatives (nine in total). 
It misses true races as the race did not manifest during execution, \eg{} in \textcode{\indigob{}}. 
Often, the hard-to-find races manifest upon specific inputs and/or event interleaving, causing dynamic techniques such as \iguard{} to miss them. 
However, it does not report false positives, as a race must manifest for \iguard{} to report. 
Overall, it reports six races and nine false negatives.

\gklee{} selectively employs symbolic execution for simple programs (\eg{} fixed kernel parameters). 
Otherwise, it falls back to dynamic analysis as in many of the programs (\eg{} \textcode{\indigoc{}}). 
Consequently, it misses several true races. 
It also reports four false positives due to the conservative nature of symbolic execution. 
Further, it cannot run many programs for various reasons, \eg{} lack of support for acquire-release (NS in~\autoref{tab:eval:main}).
Overall, \gklee{} reports nine races, four false negatives, and four false positives. 

\gverify{}, a static analysis tool, does not generate false negatives as expected. 
However, it reports many false positives. 
For example, it reports false positives for \textcode{\kaldia{}}, as it ignores asserts in the host code and the thread grid dimension in \textcode{reduce}. 
It ignores the relation among kernel parameters in \textcode{\tone{}}, bounds information in \textcode{\lud{}}, and the relation between kernel parameters and the memory allocation sizes in \textcode{\texBind{}}, generating false positives.  
It is agnostic to acquire-release synchronization (\eg{} \textcode{matMul}) and intra-warp barriers (\eg{} \textcode{warpCommunicate}). 
In total, \gverify{} reports $35$ false positives  -- $\sim$$79\%$ of the reported races!
Ignorance of semantic information available in the host code causes most of these ($24$)  

\faial{}, a static analysis tool, unfortunately reports $one$ race per kernel (if it exists) even if the kernel has multiple races (\eg{} \textcode{\indigoa{}}). 
It reports \textit{both} false negatives and positives. 
False negatives occur because it cannot detect intra-warp races (\eg{} \textcode{warpCommunicate}). 
It reports many false positives for the same reasons as \gverify{}, \ie{} agnostic of semantic information in the host code and acquire-release. 
In total, \faial{} reports $6$ false negatives and $16$ false positives. 
Most of these false positives ($14$) occur as it ignores the semantic information in the host code. 

\begin{table}
\centering
\scriptsize
\caption{Programs benefiting from host code information.}
\label{tab:eval:filter}
\begin{tabular}{l|l}\hline
   \textbf{Class of semantic information from host code}   & \textbf{Programs benefited} \\ \hline
    Asserts in host code                                  & \kaldia{}                      \\
                                                          & \kaldib{}                      \\ 
                                                          & \kaldic{}                      \\ \hline
    Thread grid dimensions set by host                    & \reduction{}                   \\ 
                                                          & \random{}                      \\ 
                                                          & \expdist{}                     \\ \hline
    Relation among kernel parameters                      & \tone{}                        \\
                                                          & \convSep{}                     \\ \hline
    Bounds on kernel parameters                           & \lud{}                         \\
                                                          & \nw{}                          \\ \hline
    Relation between allocation size and parameters       & \texBind{}                     \\
                                                          & \copyRows{}                    \\ \hline
\end{tabular}
\end{table}

\racetool{} provides the most accurate race detection. 
It detects $15$ true races but does \textit{not} report any false negatives or positives in the evaluated programs. 
\autoref{tab:eval:filter} lists the programs that benefit from each class of semantic information (\autoref{sec:motivation:main}). 
For example, \textcode{\kaldic{}} copies one matrix to another. 
Asserts ensure that rows and cols of both matrices are related. 
The kernels in \textcode{\random{}} and \textcode{\expdist{}} are always launched with one threadblock.  
The kernel in \textcode{\convSep{}} is launched with thread grid dimensions derived from the image width, a kernel parameter. 
The \textcode{\nw{}} kernel performs tiled computation, where the \textcode{tile} kernel parameter is bound by a loop iterator variable in the host code. 
The \textcode{\copyRows{}} kernel receives the size of memory allocation as a kernel parameter and uses it to access data structures. 
\racetool{} analyzes host code to generate additional constraints for the SAT solver, enabling accurate detection of races. 


\mysubsection{Overhead analysis}
\label{sec:eval:over}

\racetool{} adds \emph{no} runtime overheads since it only performs compile-time analysis. 
It needs to run only \emph{once} for a program, and if the program is modified.
The compilation took between a few milliseconds and five minutes (\textcode{\copyRows{}} took the longest).
The SAT solver accounted for most of this time, and not the analysis itself. 
However, this is a one-time cost during the debugging phase and does not affect deployment.

In contrast, \iguard{}, a dynamic tool, adds runtime overheads ranging from 1.5$\times$ to 60$\times$. 
Ideally, it should run with every execution of a program to detect data races that may manifest in that execution. 
However, high runtime overheads preclude it from production deployment. 
\gklee{} either performs symbolic execution or executes with concrete (actual) inputs.
For many programs, \eg{} \textcode{matMul}, \textcode{rule110}, it did not terminate even after a day. 
For programs where it terminates, the analysis took $1$-$43$ seconds. 
Like \racetool{}, \gverify{} and \faial{} perform compile-time analysis that adds no runtime overhead. 
\gverify{} takes $1$-$90$ seconds to compile the programs.
While \faial{} is faster -- taking only $1$-$7$ seconds, it misses many true races and also reports many false positives. 
\mysection{Related work}
\label{sec:comparison}

Many recent works have proposed techniques to detect data races in GPU code~\cite{boyer2008ada,GRace,GMRace,CUDARacecheck,LDetector,HAccRG,ScoRD,Simulee,BARRACUDA,CURD,gpuverify,gpumc,PUG,engStatTool,bardsley,dartagnan,Faial,Faialaa}.
However, these either miss true races, \ie{} false negatives~\cite{iGUARD,HiRace}, and/or generate many false positives~\cite{gpuverify,GKLEE}, and/or are designed to detect only a subset of types of race~\cite{HiRace,gpumc,gpuverify,Faialaa}, and/or need extensive modification to the program's source~\cite{gpumc,HiRace}.
\racetool{} avoids these limitations by accurately detecting a broad class of data races without false positives and needing no application modification. 

Several prior works have proposed dynamic (runtime) analysis to detect data races in GPU programs~\cite{boyer2008ada,GRace,GMRace,CUDARacecheck,LDetector,HAccRG,ScoRD,Simulee,BARRACUDA,CURD}.
Unfortunately, they all need races to manifest during testing and are thus vulnerable to missing hard-to-find critical races. 
They also incur large runtime overheads (\eg{} $60\times$ for \iguard{}~\cite{iGUARD}), precluding their from use in production. 
Several of these techniques are also oblivious to acquire-release synchronization~\cite{Simulee,BARRACUDA,CURD,LDetector} or ignore data races in global memory~\cite{boyer2008ada,GRace,GMRace,CUDARacecheck}.
A few require changes to the hardware~\cite{ScoRD,HAccRG}.
Unlike them, \racetool{} misses no races, incurs no runtime overhead, requires no hardware changes, and detects all types of races in global memory. 

Many static analysis techniques have been proposed to detect potential data races during compilation~\cite{gpuverify,gpumc,PUG,engStatTool,bardsley,dartagnan,Faial}.
These works are susceptible to reporting many false positives as they analyze only kernel code, adversely affecting their race detection accuracy. 
Unlike them, \racetool{} is the first to harness semantic information in the host code to avoid false positives. 
Furthermore, none of the prior static analyses can detect intra-warp races~\cite{dartagnan,gpumc,Faial,Faialaa}, and many are oblivious to acquire-release in GPU programs~\cite{gpuverify,PUG,engStatTool,bardsley,Faial,Faialaa}. 
\gpumc{}~\cite{gpumc} requires GPU programs to be manually converted to CPU programs since it leverages CPU model checkers~\cite{genmc,esbmc} for its analysis.
In contrast, \racetool{} accurately detects a wide range of data races while limiting/eliminating false positives and needs no programmer intervention. 

Finally, race detection has been extensively studied for multi-threaded CPU programs~\cite{benderCPU,pacer,fasttrack,ifrit,avio,eraser,CppRace,sanitizer,hard}. 
While they provide inspiration, they cannot be directly applied to GPU programs due to the massive parallelism and advanced synchronization primitives present in GPUs. 

\mysection{Conclusion}
\label{sec:conclusion}

Existing tools to detect data races for GPU-accelerated programs are not accurate, \ie{} they miss true races or report false positives. 
We observe that the host (CPU) code that launches GPU kernels contains crucial semantic information about the values that GPU kernel parameters can assume during execution. 
We build \racetool{}, which leverages this observation to accurately detect broad types of data races in GPU kernels. 


\bibliographystyle{plain}
\bibliography{sample-base}

\appendix

\section{Supporting Scoped Operations}
\label{sec:appendix:scope}

\subsection{Background on scopes}
\label{sec:appendix:background:scope}

Synchronizing across all the threads of a GPU kernel is slow and often unnecessary due to the GPU’s hierarchical programming paradigm. 
Modern GPUs thus support \textit{scopes} that enable programmers to qualify synchronization operations (\eg{} \textcode{fence}, \textcode{atomic}) with a scope qualifier. 
\cuda{} exposes three scopes to the programmer -- block, device, and system. 
The scope of an operation is the set of threads that are guaranteed to observe the effect of that operation. 
For example, block-scoped operations (\eg{} \textcode{atomicCAS\_block}) synchronize threads within the threadblock of the issuing thread.  
In contrast, device-scoped operations (\eg{} \textcode{atomicCAS}) synchronize all the threads of a GPU kernel. 
The system scope synchronizes threads across all the GPUs (\ie{} multi-GPU systems) and CPUs. 
\racetool{} focuses on programs that run on a single GPU. 
Thus, it limits the scope qualifier to block and device. 

\subsection{Extending \racetool{} with scoped operations}
\label{sec:appendix:impl:scope}



\myparagraph{Preliminary race checks \component{1}}
In this step, \racetool{} checks if the constituent \access{}es of an \accesspair{} are both atomic operations. 
It extends these checks to consider scope, which is set based on the type of race that the \accesspair{} may not participate in. 
For example, to avoid all race checks (\textbf{Trivially race-free} condition, Section 6.1 of the main paper) for an \accesspair{}, \racetool{} uses device scope. 
This checks if the constituent \access{}es are both device-scoped atomic operations. 
Further, it uses block scope to avoid intra-threadblock and intra-warp race checks for an \accesspair{} (\textbf{Avoiding intra-threadblock and intra-warp checks} condition). 
This checks if the constituent \access{}es are both either block- or device-scoped atomic operations. 

\myparagraph{Scoped acquire-release}
\racetool{} associates a scope qualifier with acquire-release synchronization operations. 
The scope of acquire or release operations is the narrower of the scopes of its constituent \textcode{atomic} and \textcode{fence} operations. 
For example, if an acquire is implemented using a block-scoped atomic and a device-scoped fence, the scope of acquire is block. 

Similarly, the scope of a lock is the narrower scope of its acquire and release operations. 
For example, if a lock has a device-scoped acquire and a block-scoped release, the scope of the lock is considered to be block. 

\racetool{} enhances the conditions for checking races in the presence of acquire-release to be aware of scopes. 
For example, to prevent an inter-threadblock conflict between an \accesspair{}, the constituting data \access{}es \textit{must} be guarded by an acquire-release with device scope. 
If \racetool{} finds a block-scoped acquire-release guarding these data \access{}es, it declares a race due to \textit{insufficient} scope. 

\myparagraph{Scoped synchronization \component{5}}
If  \racetool{} finds a conflicting \accesspair{} (Section 6.4, main paper), it then checks if the race exists in the presence of \textit{scoped} acquire-release synchronization. 
Recall that this involves synchronization, typically achieved through locks or a producer-consumer pattern. 
This step additionally receives the type of conflict found in the previous step (edge from `conflict?' to `Generate constraints for acquire-release' in Figure 6, main paper). 
The scope of acquire-release operations must be at least as wide as the conflict detected for the \accesspair{} (discussed earlier). 
\racetool{} infers this from the type of conflict detected. 
The required scope of acquire-release is device for inter-threadblock races, and at least block for intra-threadblock and intra-warp races. 
With this, \racetool{} establishes the minimum (width) scope of acquire-release required to synchronize the \accesspair{}. 

\begin{table}[t!]
    \caption{\high{Scope}-extended methods to detect races.}
    \label{tab:impl:methods:scope}
    \footnotesize
    \centering
    \begin{tabular}{c|l} \hline
       \textbf{Method}                       & \textbf{Expectation}                               \\ \hline
       syn(A, \high{scope})                  & get acquire-release for \access{} A      \\ 
                                             & that is at least as wide as scope                  \\ \hline
    \end{tabular}
\end{table}

\begin{figure}[t!]
\centering
\begin{pycode}
def getAcquire(access, scope): # acquire 
  searchForCas = false
  for op in OperationsBefore(data):
    if dominates(op, data) and isFence(op) 
       and scopeFence(op) >= scope:
       # fence found, check for CAS
      searchForCas = true
    elif searchForCas and dominates(op, data)
      and isAtomicCas(op) and scopeAtomic(op) >= scope:
      # found a fence and CAS
      return access(op)

  return NULL           # No acquire found

def syn(access, scope): # acquire/release for access
  acquire = getAcquire(data, scope)
  release = getRelease(data, scope)
  # Rest is similar to Figure 9
\end{pycode}
\caption{Getting scoped acquire/release for an \access{}.}
\label{fig:impl:lock:scope}
\end{figure}

To find the scoped acquire-release (synchronization) \access{}es, \racetool{} extends the \textit{syn} method to receive the scope argument (\autoref{tab:impl:methods:scope}). 
It returns the acquire/release operations with the required scope that guards the data \access{} (if present). 
Specifically, it checks for the presence of \textcode{atomic} and \textcode{fence} operations that are at least as wide as the required scope. 
This check ensures that \textit{syn} never returns an insufficiently scoped acquire/release operation. 

Let us trace the \textit{syn} function for an \access{} to get a device-scoped lock \access{} (\autoref{fig:impl:lock:scope}, line $15$). 
First, \racetool{} invokes the \textcode{getAcquire} method to find a device scope acquire \access{} (line $16$). 
Recall that an acquire consists of an \textcode{atomicCAS} followed by a \textcode{fence} operation~\cite{CUDAProgrammingGuide}. 
This \textit{backward} traces the kernel code's MLIR  from the data \access{} to:
First, it looks for a device-scoped \textcode{fence} that dominates the data \access{} (lines $4$-$5$). 
Next, it looks for a device-scoped \textcode{atomicCAS} that dominates the data \access{} (lines $8$-$9$), only if the required \textcode{fence} is found. 
If this pattern is found, \racetool{} returns the \access{} corresponding to the \textcode{atomicCAS} (line $11$). 
Otherwise, it returns a \textcode{NULL} \access{} (line $13$). 
Similarly, \racetool{} invokes the \textcode{getRelease} method to find the \access{} corresponding to a device-scoped release operation (not shown). 
The rest of the \textit{syn} method is similar to Figure 9 in the main paper. 
The same extensions apply to the producer-consumer pattern (not shown in the figure). 

Recall that, for a race to exist, the synchronization \access{}es corresponding to the conflicting \accesspair{} must map to different addresses (indices), \ie{} inequality. 
\racetool{} generates constraints from the corresponding \textit{scoped} synchronization \access{}es (similar to Section 6.5 of the main paper). 
For example, if the detected race is inter-threadblock, \racetool{} checks if a device-scoped synchronization \access{} is available. 
The inequality is considered trivially true if the synchronization \access{}es are not available. 
Otherwise, it generates constraints from the synchronization \access{}es. 

\end{document}